\documentclass[twocolumn]{aastex62}
\usepackage{graphicx}
\usepackage{subfigure}

\def\ie{{\it i.e. }}
\def\eg{{\it e.g. }}

\shorttitle{Bias-Free Selection Criterion}
\shortauthors{Hekun Li et al.}

\begin{document}

\title{Towards a Bias-Free Selection Criterion in Shear Measurement}

\correspondingauthor{Jun Zhang}
\email{betajzhang@sjtu.edu.cn}

\author{Hekun Li}
\affiliation{Department of Astronomy, Shanghai Jiao Tong University, Shanghai 200240, China}
\author{Jun Zhang*}
\affiliation{Department of Astronomy, Shanghai Jiao Tong University, Shanghai 200240, China}
\affiliation{Shanghai Key Laboratory for Particle Physics and Cosmology, Shanghai 200240, China}
\author{Dezi Liu}
\affiliation{South-Western Institute for Astronomy Research, Yunnan University, Kunming 650500, China}
\affiliation{The Shanghai Key Lab for Astrophysics, Shanghai Normal University, 100 Guilin Road, Shanghai 200234, China}
\author{Wentao Luo}
\affiliation{Kavli Institute for the Physics and Mathematics of the Universe, University of Tokyo, Chiba, 277-8582, Japan}
\author{Jiajun Zhang}
\affiliation{Center for Theoretical Physics of the Universe, Institute for Basic Science (IBS), Daejeon 34126, Korea}
\author{Fuyu Dong}
\affiliation{Department of Astronomy, Shanghai Jiao Tong University, Shanghai 200240, China}
\author{Zhi Shen}
\affiliation{Department of Astronomy, Shanghai Jiao Tong University, Shanghai 200240, China}
\author{Haoran Wang}
\affiliation{Department of Astronomy, Shanghai Jiao Tong University, Shanghai 200240, China}

\begin{abstract}
Sample selection is a necessary preparation for weak lensing measurement. It is well-known that selection itself may introduce bias to the measured shear signal. Using image simulation and the Fourier\_Quad shear measurement pipeline, we quantify the selection bias in various commonly used selection criteria (signal-to-noise-ratio, magnitude, etc.). We propose a new selection criterion defined in the power spectrum of the galaxy image. This new selection criterion has a low selection bias, and it is particularly convenient for shear measurement pipelines based on Fourier transformation.

\end{abstract}


\section{Introduction} \label{sec:intro}
The large foreground structure perturbs the light rays emitted from a distant galaxy, and causes a slight and coherent distortion of its shape. Such an effect is the so-called weak lensing or cosmic shear \citep{Bartelmann2001, Hoekastra2008,Kilbringer2015}. Over the years, weak lensing has become one of the most promising probes of the large cosmic structure and the expansion history of the Universe through a number of large scale weak lensing surveys such as CFHTLenS\footnote{\url{http://www.cfhtlens.org}}\citep{Heymans2013,Kilbinger2013}, KiDS\footnote{\url{http://kids.strw.leidenuniv.nl}}\citep{Hildebrandt2016}, HSC \footnote{\url{https://hsc.mtk.nao.ac.jp/ssp/survey}}\citep{Hikage2019}, and DES \footnote{\url{https://www.darkenergysurvey.org}}\citep{Troxel2018}. 

A lot of efforts have been put into the study of systematics in the measurement to match the improvement of the surveys \citep{Bridle2010, Mandelbaum2014, Mandelbaum2015}. It is challenging to obtain an unbiased shear estimator due to the systematics, including (but are not limited to) the modelling bias \citep{Bernstein2010,Voigt2010,Kacprzak2014}, the noise bias \citep{Refregier2012,Kacprzak2014}, the intrinsic alignment \citep{Troxel2015}, and the selection bias \citep{Hirata2003}. Calibrations based on the specific survey, which are useful to investigate the origins of different biases, are commonly applied to calibrate the bias for the shear measurement pipeline \citep{Kitching2008, Fenech2017}.

An interesting and important type of shear bias is caused by the imposition of the galaxy selection criteria. This happens whenever the selection criterion correlates with the galaxy shape/shear \citep{Mandelbaum2014}. The selection bias can be easily introduced in the stages of the galaxy detection and selection, and finally biases the shear measurement. For example, when the galaxies are aligned with the point spread function (PSF), they are preferentially detected because of the increasing brightness \citep{Kaiser2000, Berstein2002}. It is also pointed out that galaxies aligned orthogonally to the intrinsic shear may be preferentially selected, as existing detection algorithms favor the detections of circular objects \citep{Hirata2003}. At the faint end, it is generally difficult to clarify the influence of the detection-related selection bias, because the detection algorithms usually correlate with image properties such as the signal-to-noise ratio (SNR), galaxy size, ellipticity, and PSF profile in complicated ways \citep{Fenech2017, Liu2018, Mandelbaum2018a}. It is therefore useful to consider a cutoff on a certain image property/selection criterion, \eg, SNR or magnitude, for eliminating the selection bias arising from detection \citep{Berstein2002}.

In practice, the magnitude and the resolution factor (a way to quantify the galaxy size relative to the PSF size, see Eq.(\ref{RF})) are often used as the selection criteria \citep{Mandelbaum2013, Cardone2014} to select the relatively brighter and larger galaxies. The calibration of shear measurement for KiDS \citep{Fenech2017} shows the multiplicative bias and additive bias that are strongly magnitude-dependent, using the sample detected by both SExtractor \citep{Bertin1996} and the pipeline of KiDS. \cite{Li2018} shows that the cutoff on measured resolution factor rather than the intrinsic one biases the measurement significantly. It is becoming clear that the inappropriate selection criterion can lead to significant shear measurement bias, which is often highly nontrivial to calibrate. Removing the selection bias is therefore one of the key issues for high precision shear measurement.

In several shear measurement methods proposed recently, corrections to the selection bias have been specifically discussed, such as the Bayesian Fourier Domain (BFD) method \citep{Bernstein2014, Bernstein2016} and the Matecalibration method \citep{Huff2014, Sheldon2017}. \cite{Li2018} iterate the selection to correct the coupling between the selection criterion and shear signal, and suppress the selection bias to the subpercent level. In this paper, we focus on the Fourier\_Quad method \citep{Zhang2015, Zhang2016}, which performs the shear measurement on the power spectra of the galaxy images. We aim to propose an appropriate selection criterion, which would not induce selection bias to the measurement in the source selection stage. 

In \S\ref{shear_measure}, we briefly review the Fourier\_Quad shear measurement method. We introduce a new selection criterion in \S\ref{selection_c}, and compare its performance with other commonly used selection criteria using image simulations. We give a discussion of relevant issues in \S\ref{discussion}, and a brief conclusion in \S\ref{conclusion}.

\section{The Fourier\_Quad Method}
\label{shear_measure}

The shear estimators in the Fourier\_Quad method are defined on the 2D power spectrum of the galaxy image in Fourier space:
\begin{eqnarray}
\label{shear_estimator}
G_1&=&-\frac{1}{2}\int d^2\vec{k}(k_x^2-k_y^2)T(\vec{k})M(\vec{k})\\ \nonumber
G_2&=&-\int d^2\vec{k}k_xk_yT(\vec{k})M(\vec{k})\\ \nonumber
N&=&\int d^2\vec{k}\left[k^2-\frac{\beta^2}{2}k^4\right]T(\vec{k})M(\vec{k})
\end{eqnarray}
where $\vec{k}$ is the wave vector. $M(\vec{k})$ is the modified galaxy power spectrum properly taking into account the corrections due to the background and the Poisson noise:
\begin{eqnarray}
\label{TM}
&&M(\vec{k})=\left\vert\widetilde{f}^S(\vec{k})\right\vert^2-F^S-\left\vert\widetilde{f}^B(\vec{k})\right\vert^2+F^B\\ \nonumber
&&F^S=\frac{\int_{\vert\vec{k}\vert > k_c} d^2\vec{k}\left\vert\widetilde{f}^S(\vec{k})\right\vert^2}{\int_{\vert\vec{k}\vert > k_c} d^2\vec{k}}, \;\;\; F^B=\frac{\int_{\vert\vec{k}\vert > k_c} d^2\vec{k}\left\vert\widetilde{f}^B(\vec{k})\right\vert^2}{\int_{\vert\vec{k}\vert > k_c} d^2\vec{k}},
\end{eqnarray}
where $\widetilde{f}^S(\vec{k})$ and $\widetilde{f}^B(\vec{k})$ are the Fourier transformations of the galaxy image and a neighboring image of background noise respectively. $F^S$ and $F^B$ are estimates of the Poisson noise power spectra on the source and background images respectively. We require the critical wave number $k_c$ to be large enough for avoiding contaminations by the source power. The factor $T(\vec{k})$ in Eq.(\ref{shear_estimator}) is used to convert the form of the PSF to the isotropic Gaussian function, so that the correction of the PSF effect can be written out rigorously and model-independently. It is defined as $\left\vert\widetilde{W}_{\beta}(\vec{k})\right\vert^2/\left\vert\widetilde{W}_{PSF}(\vec{k})\right\vert^2$, \ie the ratio between the power spectrum of a 2D isotropic Gaussian function, $W_{\beta}(\vec{x})$\footnote{$W_{\beta}(\vec{x})$ is written as $(2\pi\beta^2)^{-1}\exp[-\vert\vec{x}\vert^2/(2\beta^2)]$}, and that of the original PSF, $W_{PSF}(\vec{x})$. 
To avoid singularities in the conversion, $\beta$ is required to be somewhat larger than the scale radius of the original PSF.  It has been shown in \cite{Zhang2015} that the ensemble averages of the shear estimators defined above recover the shear values to the second order in accuracy (assuming that the intrinsic galaxy images are statistically isotropic), \ie, 
\begin{equation}
\label{shear_measurement}
\frac{\left\langle  G_1\right\rangle }{\left\langle  N\right\rangle }=g_1+O(g_{1,2}^3),\;\;\;\frac{\left\langle  G_2\right\rangle }{\left\langle  N\right\rangle }=g_2+O(g_{1,2}^3).
\end{equation}
Note that the ensemble averages are taken for $G_1$, $G_2$, and $N$ separately \citep{ZK2011}. \cite{Zhang2016} offers another way of measuring the lensing statistics using the probability distribution function (PDF) of the shear estimators.

An appealing feature of Fourier\_Quad is its good behavior for sources at the faint/small end, \ie, the inclusion of barely-resolved galaxies or even point sources does not bias the shear measurement according to Eq.(\ref{shear_measurement}) \citep{Zhang2015}. Therefore, there are no constraints on the selection criteria that are imposed by Fourier\_Quad, making our discussion of the selection effects quite easy and straightforward. Note that this is not the case in many other shear recovery methods, in which galaxies are typically required to be resolved to a certain level for shear measurement. 

\section{Bias-Free Selection Criterion}
\label{selection_c}

Traditionally, galaxies are selected according to, \eg, the magnitude, SNR, resolution factor. The shear bias is typically calibrated as a function of these parameters \citep{Fenech2017, Liu2018, Mandelbaum2018a}. It is important to note that in these studies, there are two kinds of bias involved: one is due to the shear measurement method itself, and another is due to the selection criterion. The second type of bias is caused by the coupling between the selection criterion and the galaxy shape. An ideal selection criterion would be least coupled with the underlying shear signal. 

According to the lensing formalism, the total flux of a galaxy is only affected by the convergence, not the shear. It is therefore a promising selection criterion candidate. In practice, however, it is difficult to measure the total flux on a noisy image without being influenced by the galaxy shape. We consider a direct measure of the total flux in Fourier space using the power spectrum of the galaxy image at $k=0$. We define our new selection criterion as:
\begin{equation}
\nu_F = \frac{|\widetilde{f}^S(k=0)|}{\sqrt{N} \sigma}
\end{equation}
with $N$ and $\sigma$ being the total number of pixels in the galaxy stamp and the root mean square (RMS) of the background noise respectively. The numerator $|\widetilde{f}^S(k=0)|$\footnote{The sky background has been subtracted from the real space source image, $f^S(\vec{x})$.}, according to its definition, is the total flux within the galaxy stamp. This quantity can be measured quickly without morphological fitting or image convolution. We therefore expect it to be the least sensitive to the galaxy shape. 

In the rest of the section, we compare the performance of $\nu_F$ with several other selection criteria, including SNR, MAG\_AUTO, and the resolution factor \citep{Hirata2003,Massey2013,Mandelbaum2018b}. The SNR and MAG\_AUTO are commonly measured by SExtractor \citep{Massey2007, Leauthaud2007, Schrabback2010, Zuntz2017, Liu2018, Mandelbaum2018a}. The resulotion factor, $R_F$, is defined as the ratio between the quadrupole of PSF, $T_{P}$, and that of the galaxy, $T_{G}$. \citep{Mandelbaum2018a}. 
\begin{equation}
\label{RF}
R_{F} = 1 - \frac{T_{P}}{T_{G}}.
\end{equation}
The image quadrupole is defined as
\begin{equation}
T = \frac{\int\int (x^2 + y^2)w(x,y)I(x,y)dxdy}{\int\int w(x,y)I(x,y)dxdy},
\end{equation}
where $I(x,y)$ is the galaxy brightness distribution, and $w(x,y)$ is the weighting kernel. Because a constant PSF is used in our image simulations, we regard the quadrupole of galaxy as the resolution factor hereafter. A Gaussian weight is used in the quadrupole calculation \footnote{The scale radius of the Gaussian weight is given by the effective radius obtained from $N_{pix}=\pi r_{eff}^2$, where $N_{pix}$ is the pixel number of the source generated by SExtractor.}. We adopt the commonly used method, the multiplicative ($m_{1/2}$) and additive bias ($c_{1/2}$), to quantify the precision of shear measurement.
\begin{equation}
g^{meausre}_{1/2} = (1+m_{1/2})g^{true}_{1/2} +c_{1/2}.
\end{equation}

\subsection{Galaxy Simulations}
\label{tests}

\begin{figure}
	\centering
	\includegraphics[width=0.5\textwidth]{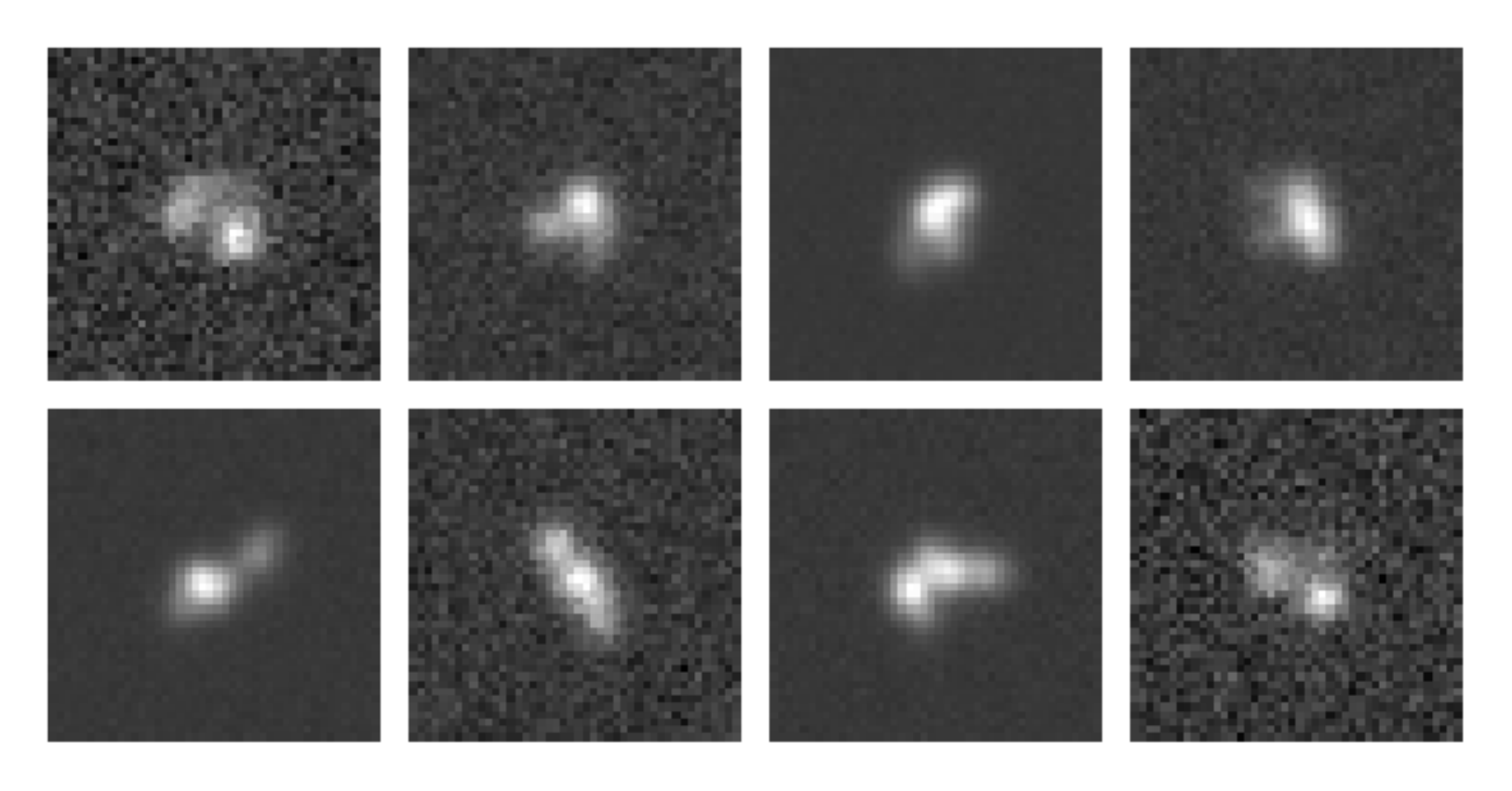}\\
	\includegraphics[width=0.5\textwidth]{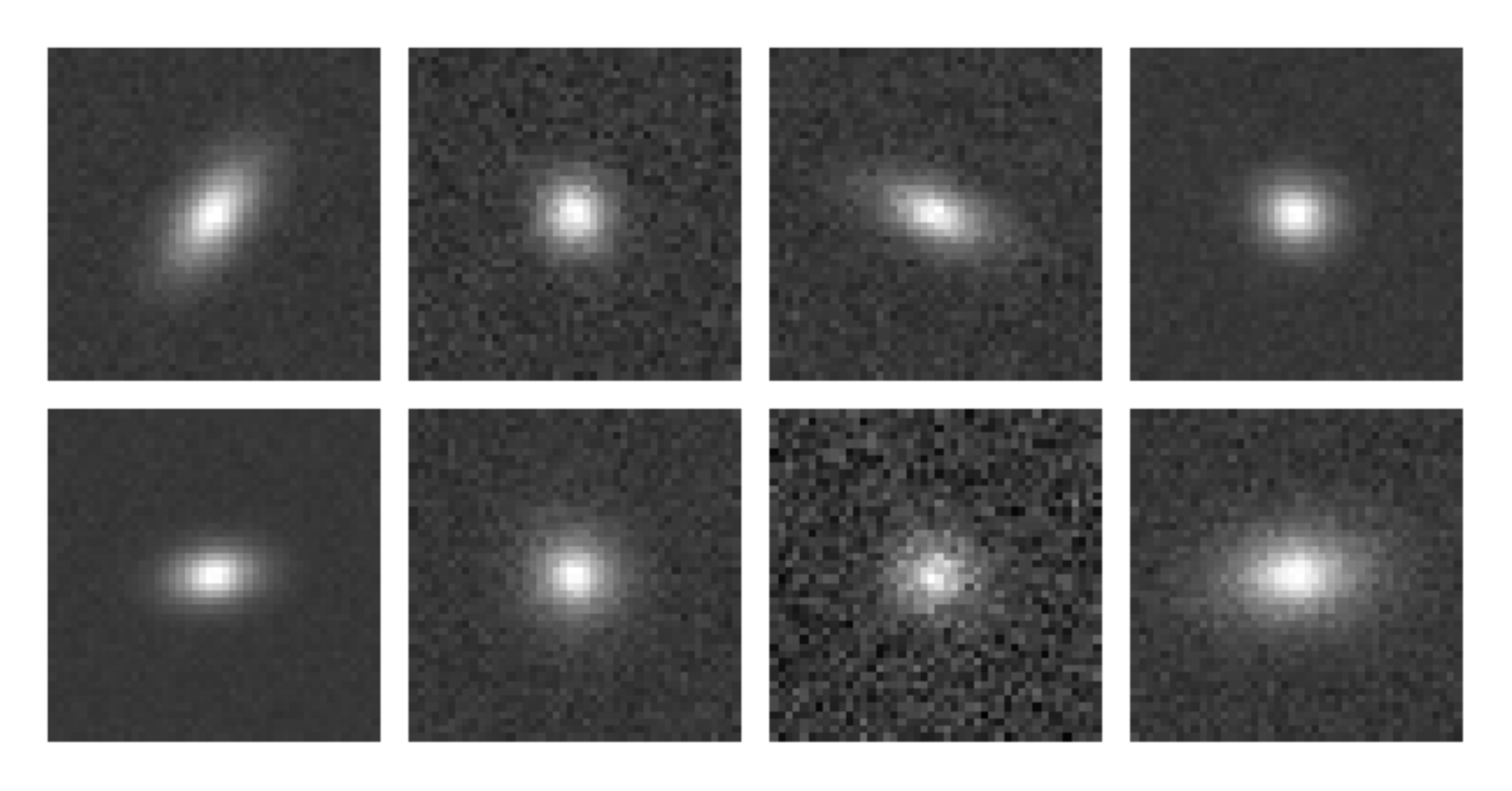}\\
	\caption{The upper two rows of panels show the point-source galaxies generated by the random-walk method to mimic irregular galaxies in real observations. The lower two rows show examples of regular galaxies generated by Galsim.  \label{fig:galaxy}}
\end{figure}

We set up two types of simulations: one uses the irregular galaxies made of point sources, whose positions are generated by random walks \citep{Zhang2008}; the other one uses parameterised regular galaxies generated by Galsim \citep{rowe2015}, an open source simulation toolkit. The parameters regarding the observational conditions are from CFHTLenS\citep{Miller2013}. We assume that the observation is made in the i-band ($i_{814}$), with each exposure time being 600 seconds, and the gain being 1.5 $e$-/ADU, the zero point being 25.77 $mag$. The pixel size is $0.187^{''}$. The stand deviation of the background noise is 60 ADUs, which is obtained by the least-squares fitting to the CFHTLenS images. We choose the galaxy stamp size to be 64$\times$64 pixels.
Figure \ref{fig:galaxy} shows some bright examples of both types of galaxies. 

\subsubsection{Point-Source Galaxies}

In lensing surveys, many sources have irregular shapes. The irregularity mainly arises from galaxy mergers. The fraction of irregular galaxy generally increases with the depth of the survey because of the increasing galaxy merger rate at high redshifts \citep{Bridge2007, Conselice2008}. Therefore, it is important to have a way modeling the morphology of irregular galaxies in shear measurement tests.

Real galaxies consist of hundreds of millions of stars which are nothing but point sources. Our irregular galaxies are made of a number of point sources using the random walk method \citep{Zhang2008}. In forming one such galaxy, \ie to determine the positions of its point sources, we let the random point walk start from the center of the stamp, and wanders for 30 steps with a fixed step size (equals one pixel) and random directions. We confine the points to a circular area of radius equals to $7$ pixels. Steps that are about to go out of the circular region are pulled back to the stamp center to continue from there. Galaxy luminosity is modeled by changing the fluxes of the point sources. Noises are added directly to the pixels to mimic different background brightness. The image generation method based on point sources has the merit of running fast, precise shape distortion, and efficient PSF convolution. The resulting irregular galaxy profiles enable model-independent studies of systematic errors in shear measurement.

\subsubsection{Galsim Galaxies}

\begin{figure}
	\centering
	\includegraphics[width=0.4\textwidth]{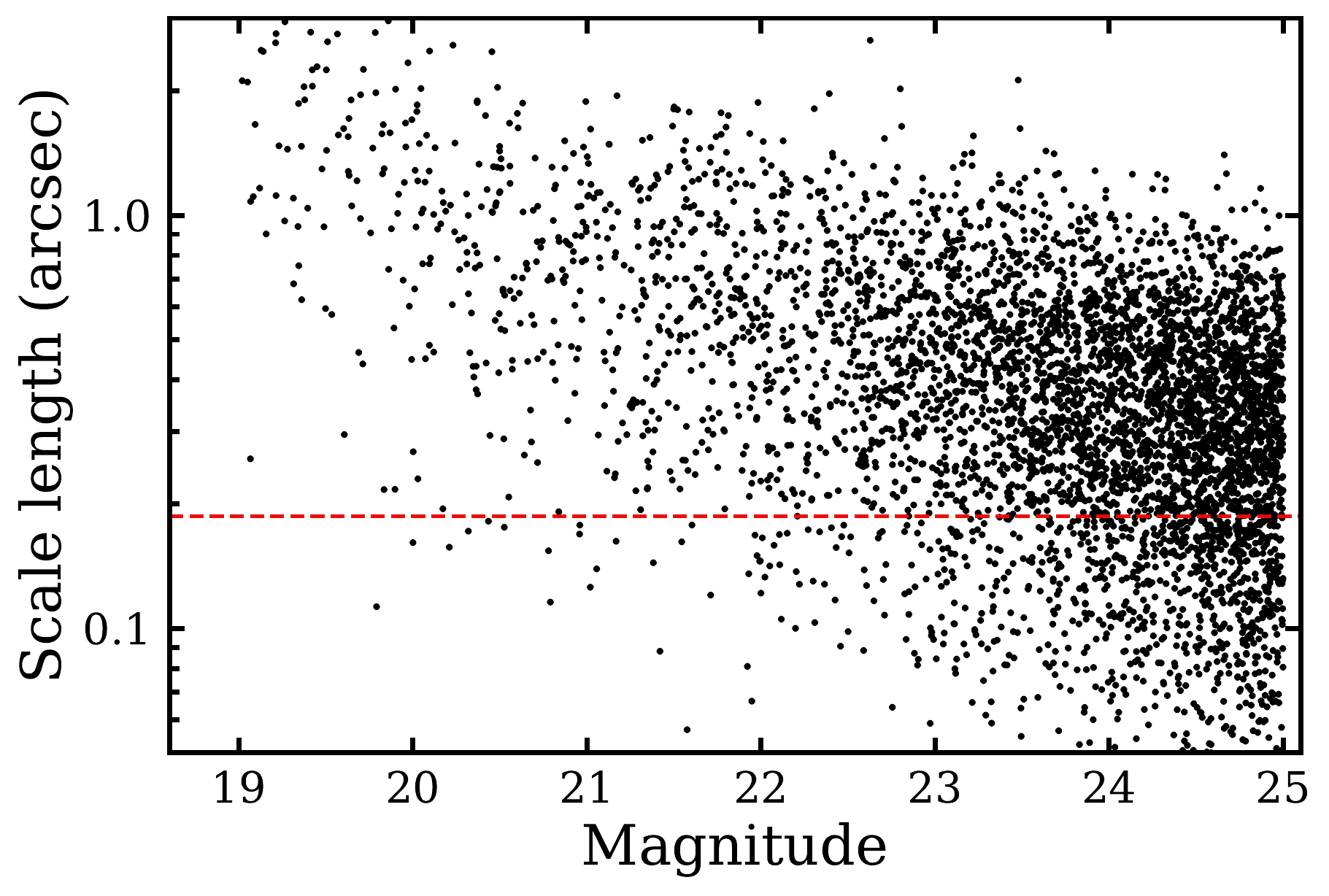}
	\caption{The scale length distribution with respect to the magnitude. The red dashed line shows the pixel scale.\label{fig:mag_ra}}
\end{figure}

Galaxies of regular morphologies are modelled with two types of profiles \citep{simar2011}: the deVaucouleurs profile (S\'{e}rsic index $n=4$) and the exponential profile (S\'{e}rsic index $n=1$). We follow \citet{Miller2013} to set up our simulations. Our galaxy sample consists of 90\% disc-dominated galaxies and 10\% bulge-dominated ones. For the bulge-dominated galaxies, the pure deVaucouleures profile is used. For the disc-dominated galaxies, the bulge-to-all fraction, $f=B/T$, is assumed to be a truncated normal distribution centered at $f=0$ with $\sigma_f=0.1$. The half-light radius of the bulge is set to be equal to the scale length of the disc, and the whole profile is truncated at 4.5 times the disc scale length to avoid the prohibitive computational cost. The distribution of the disc scale length takes the following form:
\begin{equation}
P(r)\propto r\exp\left[-(r/a)^\alpha\right]
\end{equation}
where $a=r_s/0.833$ \footnote{The factor $0.833$ is from Appendix B1 of \cite{Miller2013}, which is indeed a typo stressed by \cite{Fenech2017}. Its true value should be $1.13$. However, we still use the old value here, as it would not change our main conclusion.} and $\alpha=4/3$. $r_s$ (arcsec) is related to the $i-$band magnitude through: $\ln(r_s) = -1.145 - 0.269\times(i_{814} - 23)$. Figure \ref{fig:mag_ra} shows the mock scale length distribution with respect to the magnitude. 

We adopt different ellipticity probability distribution functions (PDF) for the disc-dominated and the bulge-dominated galaxies. For the former, we assume:
\begin{equation}	
P(e)=\frac{Ae\left[1-\exp\left(\frac{e-e_{max}}{a}\right)\right]}{(1+e)(e^2+e_0^2)^{1/2}}
\end{equation} 
with $e_{max} = 0.804$, $e_0 = 0.0256$, $a = 0.2539$ and $A = 2.4318$. It comes from the fitting to the 66762 SDSS disc-dominated galaxies from DR7 \citep{Abazajian2009}. For the bulge-dominated galaxies, we use:
\begin{equation}
P(e) = A e\exp(-\alpha e-\beta e^2)
\end{equation} 
where $\alpha=2.368$, $\beta=6.691$ and $A=27.8366$.

\subsubsection{Other Simulation Parameters}
\begin{figure}
\centering
\includegraphics[width=0.4\textwidth]{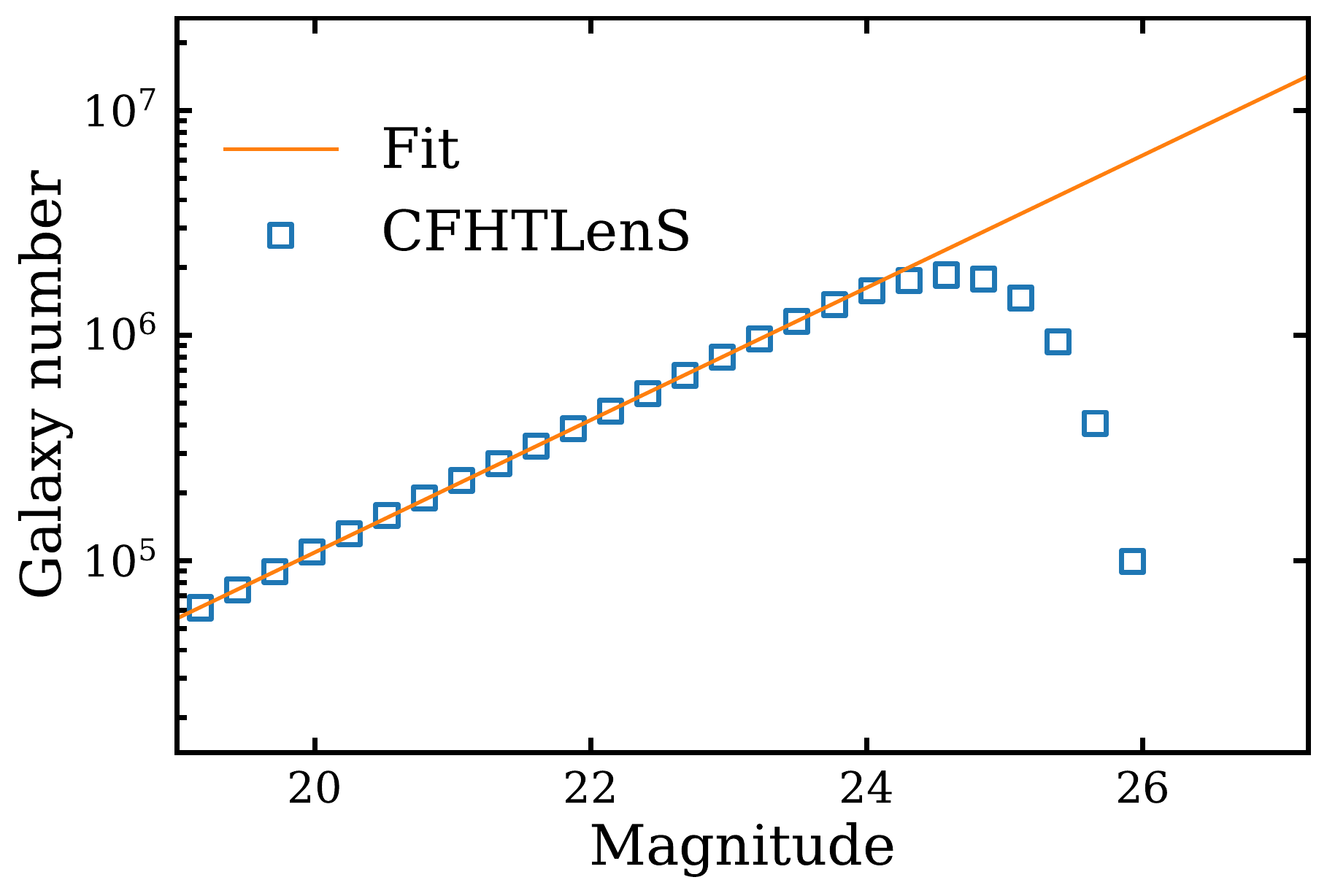}
\caption{The blue squares present the distribution of the magnitudes in the CFHTLenS catalog. The orange line is the best fitting curve from those with magnitude $<24$.\label{fig:mag_fit}}
\end{figure}

We generate two sets of samples to investigate the selection effect: the bright sample (PI for point-source galaxies, and GI for Galsim galaxies hereafter) contains the sources with magnitudes range from 20 to 24.8; the faint sample (PII for point source and GII for Galsim galaxies hereafter) consists of galaxies with magnitudes range from 23 to 24.8. The faint samples, GII and PII, are not the sub-samples cut from the bright ones.  The magnitude distribution is obtained by fitting to the CFHTLenS $i$-band catalog \citep{Erben2013} up to $24$ in $mag$, as shown in Figure \ref{fig:mag_fit}. The fitting function is extrapolated to higher magnitude for our purpose.

For the PSF, we adopt the Moffat form \citep{Bridle2009}:
\begin{equation}
I(r) \propto \left[1+\left(\frac{r}{r_d}\right)^2\right]^{-3.5}H(r_c-r).
\end{equation}
in which $r_d$ is the scale length, $r_c$ is set to 4 times $r_d$, and $H(r_c-r)$ is the Heaviside step function. 

\subsection{Shear Sensitivity \& Multiplicative Selection Bias}\label{sec:m}

The selection bias is caused by the correlation between the selection criterion and the galaxy shape. To understand this, one can parameterize the selection criterion $s$ as:
\begin{equation}\label{couple1}
s \approx s^I + \alpha g,
\end{equation}
in which $s^I$ denotes the intrinsic (pre-lensing) value of $s$, and $g$ is the underlying shear signal. Note that for simplicity, $g$ here can stand for either $g_1$ or $g_2$. $\alpha$ is the shear sensitivity coefficient. Let us assume that the galaxy ellipticity $e$ is an unbiased shear estimator, \ie, $e=e^I + g$, where $e^I$ is the intrinsic value of $e$. Applying a cut of the sample according to $s\geq s_c$, we can write down the measured shear (without weighting) as:
\begin{equation}
\hat{g} = \frac{\int_{-\infty}^{\infty}de\int_{s_c}^{\infty}ds \cdot P(e,s)\cdot e}{\int_{-\infty}^{\infty}de\int_{s_c}^{\infty}ds \cdot P(e,s)}.
\end{equation}
$P(e,s)$ is the joint probability distribution function of the ellipticity and the selection criterion. The conservation of galaxy number implies that $P^I(e^I, s^I)de^Ids^I = P(e,s)deds$. To the first order of $g$, we obtain:
\begin{eqnarray}
\hat{g} &=& \frac{\int_{-\infty}^{\infty}de^I\int_{s_c-\alpha g}^{\infty}ds^I\cdot P^I(e^I,s^I)\cdot (e^I+g)} 
{\int_{-\infty}^{\infty}de^I \int_{s_c-\alpha g}^{\infty}ds^I \cdot P^I(e^I,s^I)} \nonumber\\
&\approx& \frac{\int_{-\infty}^{\infty} de^I [\int_{s_c}^{\infty} g P^I(e^I,s^I)ds^I + \alpha g e^I P^I(e^I,s_c)]}
{\int_{-\infty}^{\infty} de^I \int_{s_c}^{\infty} P^I(e^I,s^I)ds^I} \nonumber\\
&\approx& g\cdot\left(1+\frac{\int_{-\infty}^{\infty} de^I \cdot\alpha\cdot e^I\cdot P^I(e^I,s_c) }{\int_{-\infty}^{\infty}de^I\int_{s_c}^\infty P^I(e^I,s^I)ds^I}\right).
\end{eqnarray}
Note that $\alpha$ is not pulled out of the integration, as it is usually not a constant, but a function of the galaxy properties, \eg,  $\alpha(e^I,s^I)$. It is clear that a shear-sensitive selection criterion implies a shear measurement bias if it is used to select the galaxy sample. Therefore, as a first step in studying the selection bias, we can simply observe how the selection criterion varies with the underlying shear.

\begin{figure*}
	\centering
	%\subfigure{\includegraphics[width=0.98\linewidth]{change_galsim.pdf}}
	%\includegraphics[width=0.5\textwidth]{change_pts.pdf}
	%\includegraphics[width=0.5\textwidth]{change_galsim.pdf}
	\gridline{\fig{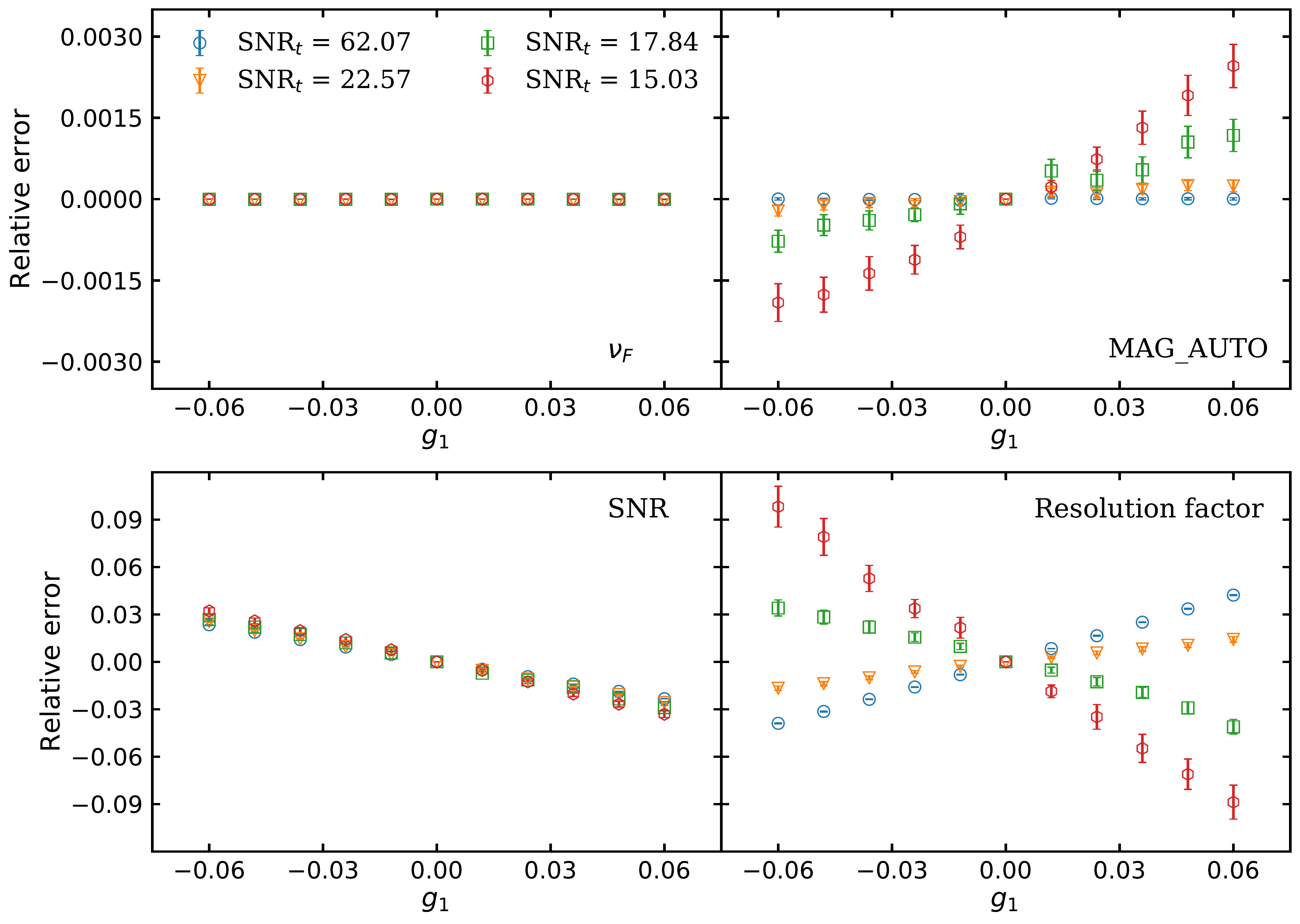}{0.47\textwidth}{Random-walk galaxy}
		\fig{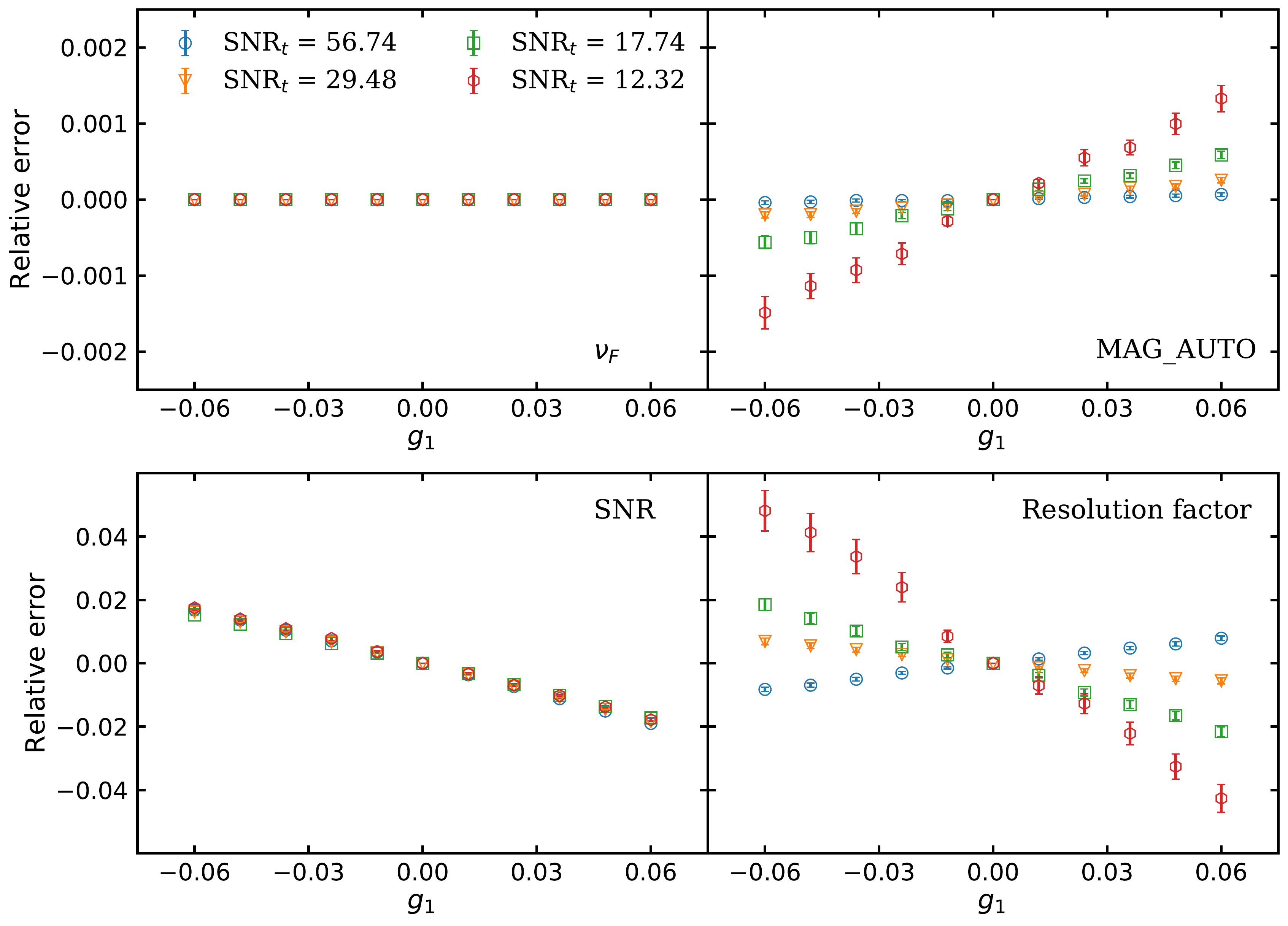}{0.47\textwidth}{Regular galaxy ($e_1=0.7,e_2=0$)}}
	\caption{The shear sensitivities of different selection criteria measured under different noise conditions. The vertical axes are called ``relative error'', referring to the relative change of the selection criterion as a function of the underlying shear. Results in the left and right panels use random-walk galaxies and Galsim galaxies respectively, both of which are chosen to be somewhat elliptical along the x axis as their intrinsic shape. The shear sensitivities in most of the plots are therefore quite significant with respect to the $g_1$ component.
		\label{fig:change}}
\end{figure*}

In figure \ref{fig:change}, we show the shear sensitivity for several different selection criteria from galaxy images of different intrinsic SNR (pre-lensing). The vertical axes in the plots are called "relative error", referring to the relative change of the selection criterion as a function of the underlying shear. The results in the left and right panels are from a single random-walk galaxy and a Galsim galaxy respectively. Different total fluxes are assigned to the galaxy to form images of specified intrinsic SNRs. To suppress the fluctuation due to noise, each data point is averaged over $200$ noise realizations. The figure shows that our new selection criterion $\nu_F$ is least sensitive to shear. There is no visible variation of $\nu_F$ even for sources with SNR $\sim 10$. MAG\_AUTO is the next best one, but a certain level of shear-sensitivity is found when SNR$\lesssim 20$. On the other hand, selection criteria such as SNR and the resolution factor are found to be strongly correlated with the galaxy shape in the figure, implying a potentially large selection bias. 

\begin{figure}
	\centering
	\includegraphics[width=0.75\linewidth]{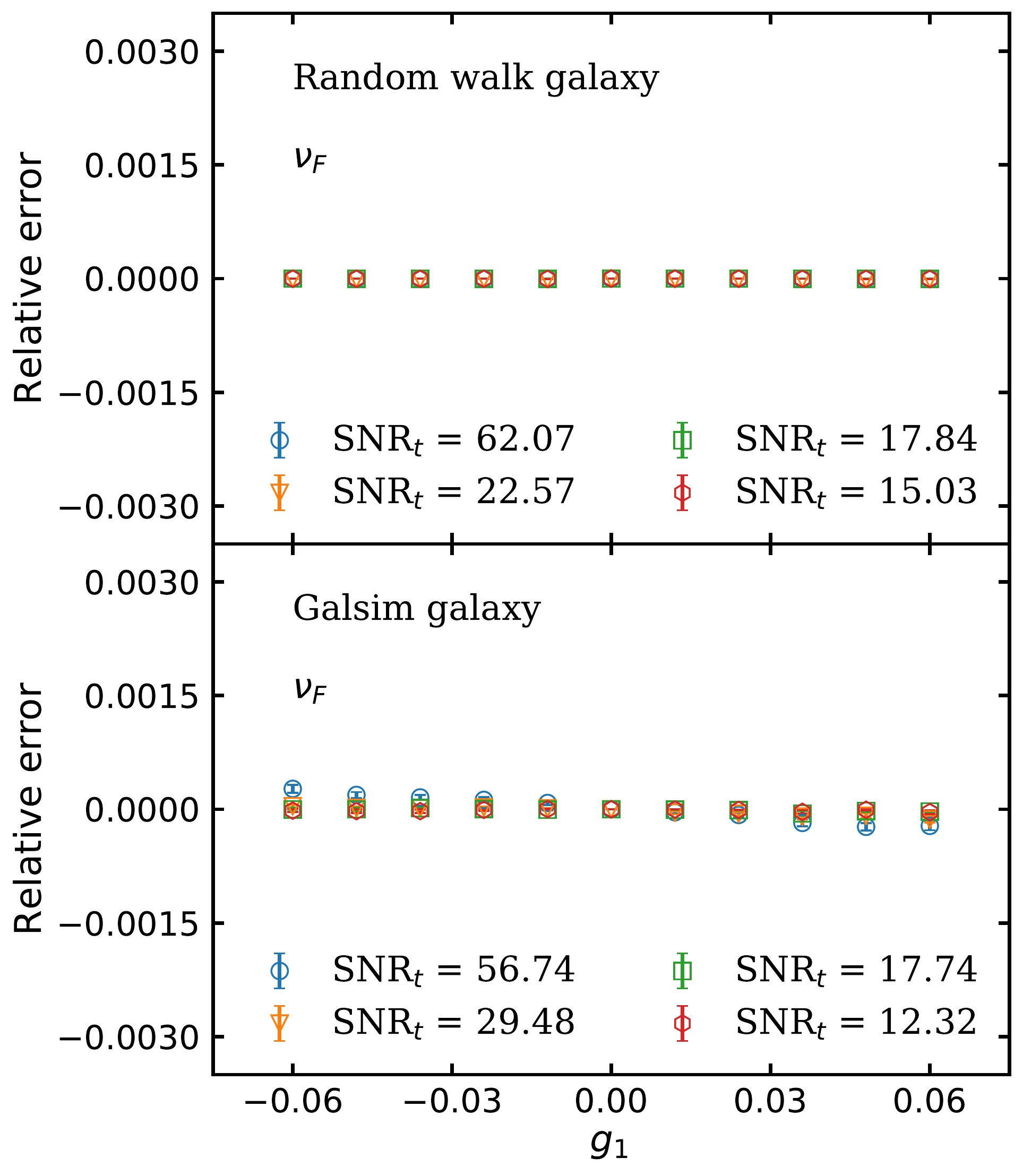}
	\caption{The result of $\nu_F$ in Figure \ref{fig:change}, reproduced with the RMS of background noise estimated using SExtractor.}\label{fig:sigma_compare}
\end{figure}

Note that in Figure \ref{fig:change}, $\nu_F$ is calculated using the input value of $\sigma$, the RMS of the background noise. We find that if $\sigma$ is estimated using SExtractor, the results would be slightly different for the brightest Galsim galaxies, as shown in Figure \ref{fig:sigma_compare}. This is likely due to the influence of the outspread profiles of bright galaxies on the estimation of $\sigma$ in SExtractor, a problem that can be avoided in principle (\eg, by using other pipelines).  Nevertheless, the shape dependence of $\nu_F$ is still much milder than those of the other selection criteria. We therefore consider $\nu_F$ as a promising candidate of selection criteria.

\begin{figure*}[htb]
\centering
\includegraphics[width=0.65\linewidth]{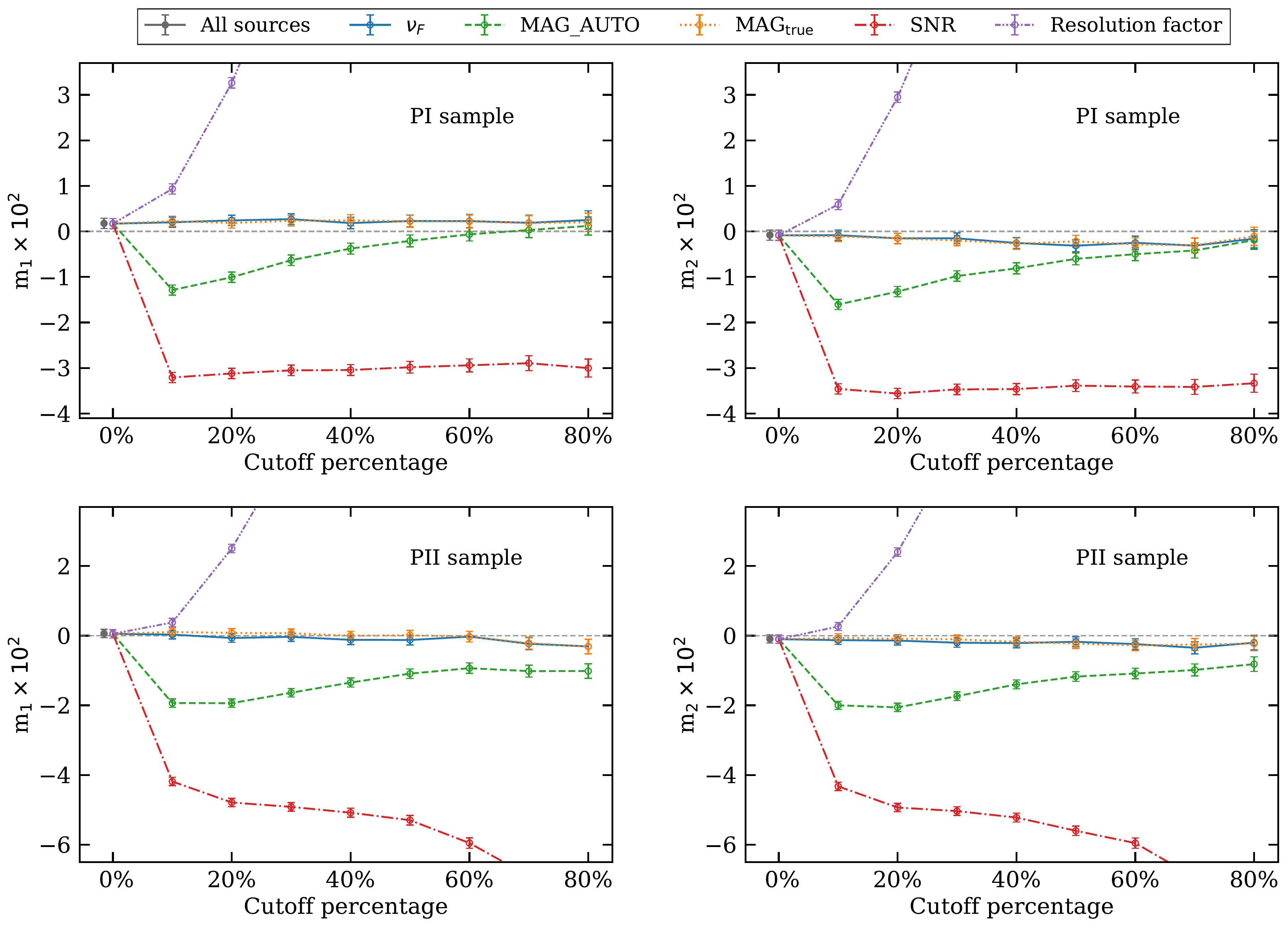}
\caption{The multiplicative biases due to the cutoffs to the irregular galaxy sample according to different selection criteria. The curves of MAG$_{\rm{true}}$(orange) should be a reference as it does not cause any selection bias. The gray solid points are the results from the entire sample (including those that are not detected by SExtractor), showing that the detection bias is insignificant. \label{fig:ptsmc}}
\end{figure*}

\begin{figure*}[htb]
\centering
\includegraphics[width=0.65\linewidth]{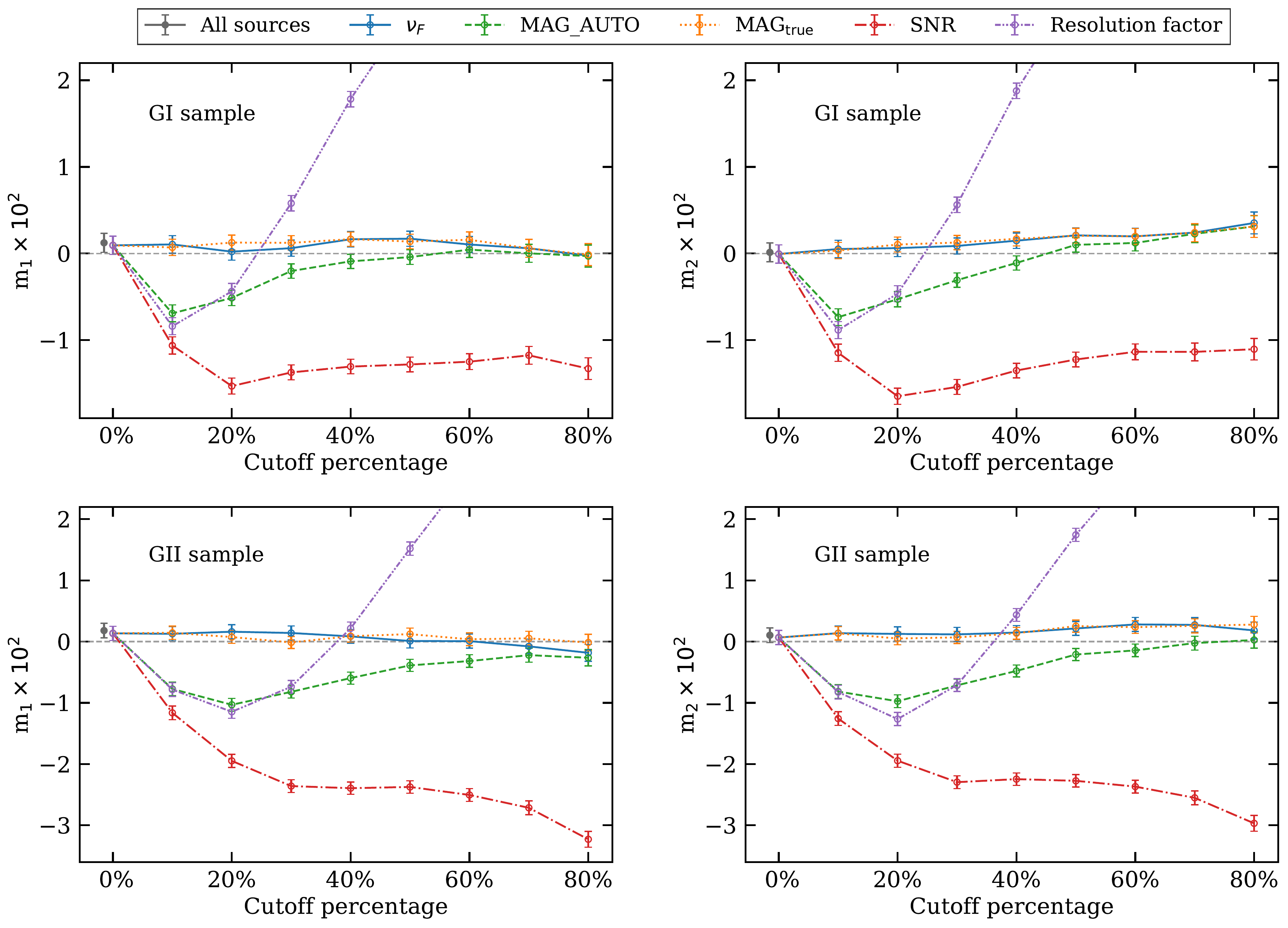}
\caption{Same as Fig.\ref{fig:ptsmc}, but for the galaxy sample of regular morphology generated by Galsim. \label{fig:galmc}}
\end{figure*}

We present the main results for multiplicative bias in Figure \ref{fig:ptsmc} and Figure \ref{fig:galmc} for point-source galaxies and Galsim galaxies respectively. To find the shear biases, we run simulations with randomly chosen shear values $g_1$ and $g_2$ in between $-0.04$ and $0.04$. We generate $1.0\times10^7$ galaxies for each set of shear values, and use $1.4\times10^8$ galaxies in total. To present the bias due to selection, we abandon the $10\%$ faintest ones of the total sample each time according to the selection criterion of our interest, until there are only $20\%$ sources left. In the figures, we also show results from cutting the galaxy samples with their intrinsic magnitudes (pre-lensing, MAG$_{\rm{true}}$) which does not introduce any selection bias. 

We find significant selection biases for SNR and the resolution factor using both regular and irregular galaxy samples, and mild ones for MAG\_AUTO. In contrast, $\nu_F$ performs consistently well, and as well as MAG$_{\rm{true}}$ indeed. These behaviors of the selection criteria are consistent with Figure \ref{fig:change}. Compared with the regular galaxy sample, the selection bias would be exacerbated by the irregularity of the galaxy morphology. The increasing portion of faint galaxies also enlarges this bias. 

Essentially, $\nu_F$ should be equivalent to magnitude as they are both  measures of the total flux. However, magnitude is typically estimated within a domain that is dependent on the galaxy shape, while the measurement of $\nu_F$ does not involve morphological constraints. It is therefore a better selection criterion than MAG\_AUTO. 

Note that in all the cases shown in Figure \ref{fig:ptsmc} \& \ref{fig:galmc}, the shear biases at zero-percentage cutoff are negligible. This fact simply demonstrates the accuracy of our shear measurement method in the absence of the selection effect (samples are almost complete, see Section \ref{discussion} for more discussions). In real observations, to guarantee the accuracy of the shear measurement, one has to select the galaxy sample at the faint end by applying a certain cut, for which our above discussions become relevant.

\subsection{Additive Selection Bias}\label{sec:a}

The selection bias can also take an additive form when the PSF is anisotropic. Let us follow similar calculations as in \S\ref{sec:m}, but this time consider the selection criterion $s$ that is not only affected by the underlying shear, but also by the ellipticity of the PSF $e^*$:
\begin{equation}\label{couple2}
s \approx s^I + \alpha g+\beta e^*
\end{equation}
Again, without loss of generality, we do not specify the subindices of $g$ and $e^*$ here. $\beta$ is the shear sensitivity of $s$ with respect to $e^*$. Let us still assume that the galaxy ellipticity $e$ is an unbiased shear estimator, \ie, $e=e^I + g$, \ie, the influence of PSF on the shear estimator is removed. We can write down the measured shear as:
\begin{eqnarray}
&&\hat{g} = \frac{\int_{-\infty}^{\infty}de\int_{s_c}^{\infty}ds \cdot P(e,s)\cdot e}{\int_{-\infty}^{\infty}de\int_{s_c}^{\infty}ds \cdot P(e,s)} \nonumber \\
&=& \frac{\int_{-\infty}^{\infty}de^I\int_{s_c-\alpha g-\beta e^*}^{\infty}ds^I\cdot P^I(e^I,s^I)\cdot (e^I+g)} 
{\int_{-\infty}^{\infty}de^I \int_{s_c-\alpha g-\beta e^*}^{\infty}ds^I \cdot P^I(e^I,s^I)} 
\end{eqnarray}
Keeping terms up to the first orders in $g$ and $e^*$, we have:
\begin{eqnarray}
\hat{g}&\approx& g\cdot\left(1+\frac{\int_{-\infty}^{\infty} de^I \cdot\alpha\cdot e^I\cdot P^I(e^I,s_c) }{\int_{-\infty}^{\infty}de^I\int_{s_c}^\infty P^I(e^I,s^I)ds^I}\right) \nonumber \\ 
&+&e^*\cdot\frac{\int_{-\infty}^{\infty} de^I \cdot\beta\cdot e^I\cdot P^I(e^I,s_c) }{\int_{-\infty}^{\infty}de^I\int_{s_c}^\infty P^I(e^I,s^I)ds^I}.
\label{additive_bias} 
\end{eqnarray}

\begin{figure*}[htb]
\centering
\includegraphics[width=0.9\linewidth]{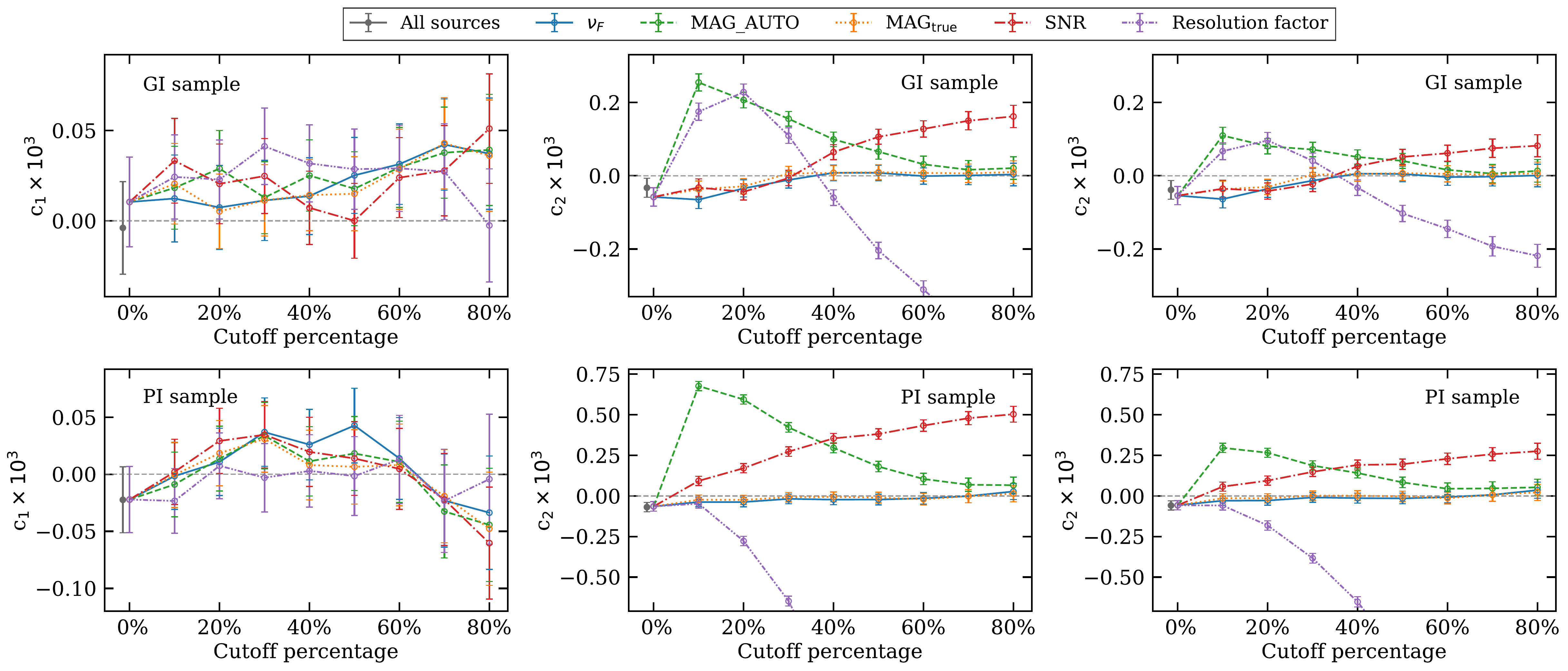}
\caption{The additive selection bias due to the existence of an anisotropic PSF for the regular galaxy sample (upper panels) and the irregular one (lower panels). The PSF has the ellipticity of $\bold{e}=(0,0.1)$ for the results in the first two columns, $\bold{e}=(0,0.05)$ for those in the last column (only $c_2$ is shown). The grey solid points are the results from the entire sample (including those that are not detected by SExtractor), showing that the detection bias is insignificant. \label{fig:epsfmc}}
\end{figure*}

The above calculation shows that one should also expect an additive shear bias if the selection criterion is correlated with the galaxy shape. This is demonstrated in Figure \ref{fig:epsfmc}, in which we shear the PSF used in the GI- and PI-sample slightly ($e_1=0,e_2=0.1, 0.05$) and repeat the selection processes, as done in \S\ref{sec:m}. According to the figure, there are significant additive biases ($c_2$ due to the non-zero $e_2$ of the PSF) in the sample selected by SNR, resolution factor, and MAG\_AUTO. The amplitude of $c_2$ is roughly proportional to $e_2$ of the PSF, as predicted in Eq.(\ref{additive_bias}). Note that when the cutoff on MAG\_AUTO comes to the bright end, the additive bias $c_2$ becomes negligible. However, the bias from SNR or resolution factor becomes significant on the bright end. On the other hand, $\nu_F$ does not seem to introduce any noticable additive bias for either faint or bright sources.

\section{Discussions}
\label{discussion}
We discuss some details relate to the results above, including detection bias, scatter of $\nu_F$ due to noise, and using $\nu_F$ to weight the shear estimators in Fourier\_Quad method.

\subsection{Detection bias}\label{sec:detection}

Source detection is a necessary step in practice. At the faint end, typically, whether a galaxy is detected or not depends on its brightness, morphology, as well as PSF in a complicated way. This pre-selection step causes a systematic change of the morphological distribution of the galaxies at the faint end, therefore can naturally introduce a shear bias. This is what we call the detection bias. 

The detection bias is mixed with the selection effect at the very faint end. To suppress the detection bias (as it is not the focus of this work), we set a low threshold in SExtractor to include more sources. We require at least 5 pixels above $1.5\sigma$ of the background noise for a detection. The detection rate is about $95\%$ in the Galsim samples (both GI and GII), and $99.9\%$ for the random-walk samples (both PI and PII). In every case, we find that the multiplicative and additive bias of the entire sample, shown as grey circles in Figure \ref{fig:ptsmc}, \ref{fig:galmc}, and \ref{fig:epsfmc}, are very close to that of the detected sample (non-cut sample). Therefore, the detection bias problem is not important in our current simulations. On the other hand, for real data, a high cut-off on the galaxy sample in terms of the selection criterion should always allow us to avoid the detection bias.

\subsection{Reducing the scatter of $\nu_F$}\label{sec:scatter}

\begin{figure}
	\centering
	\includegraphics[width=0.4\textwidth]{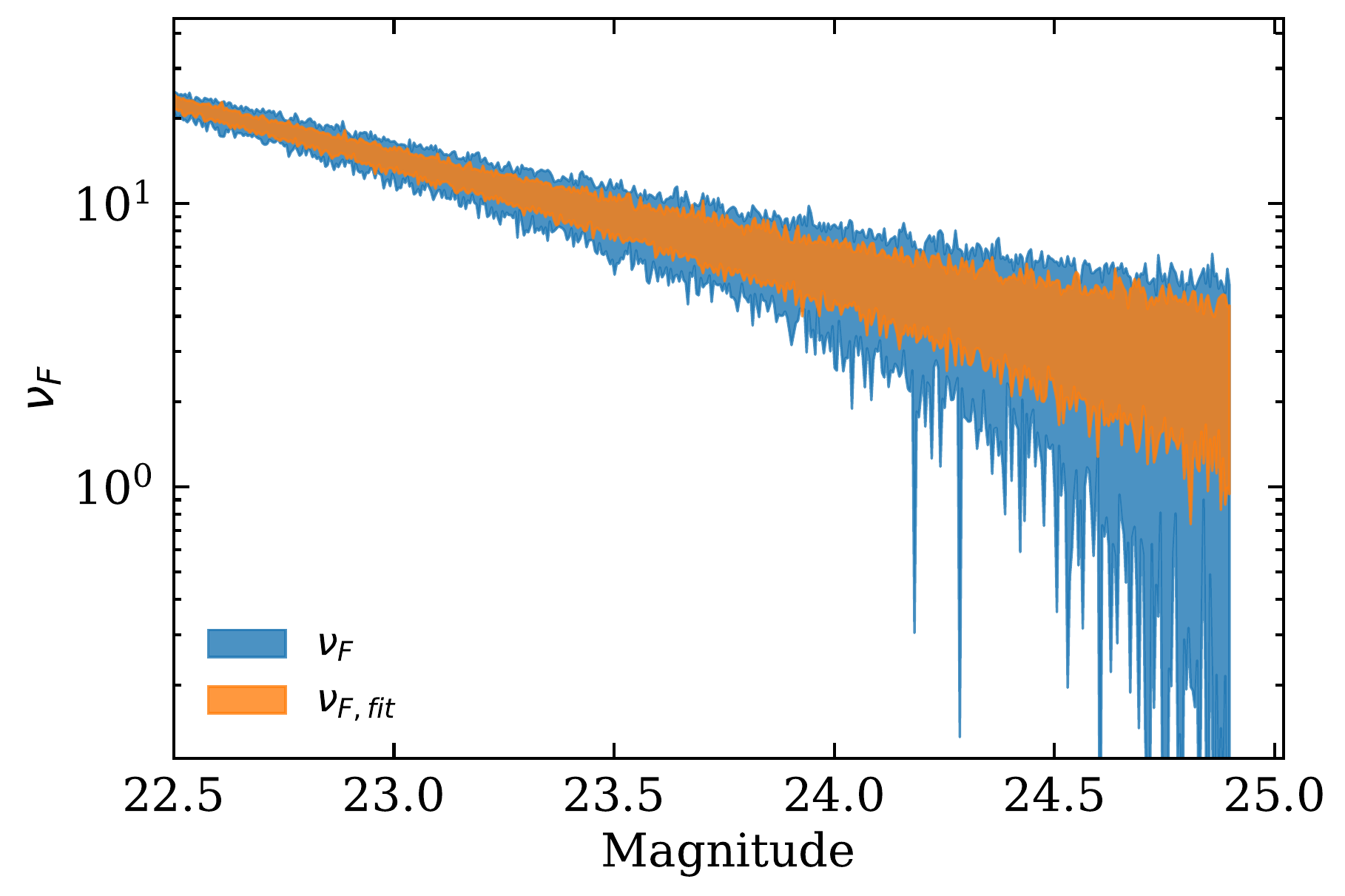}
	\caption{Scatters of $\nu_F$ (in blue) and $\nu_{F,fit}$ (in orange). \label{fig:scatter}}
\end{figure}

According to its definition, $\nu_F$ is the total flux of the galaxy in the noise-free image. However, the existence of the background noise can significantly scatter the value of $\nu_F$, particularly at the faint end, as shown in Figure \ref{fig:scatter} with blue color. To reduce the scatter of $\nu_F$, we fit a 2nd-order polynomial function in the neighborhood of $k=0$ in Fourier space to recover the $\nu_F$. We use the neighboring 5$\times$5 areas to fit the logarithms of the pixel values around $k=0$. The logarithmic scale makes the profile more smooth. The pixel values at $k=0$ and the four corners of the 5$\times$5 region are excluded from the fitting to make the fitting region more isotropic. As a result, the $\nu_F$ derived from fitting (called $\nu_{F,fit}$ hereafter) is much less scattered, as shown in Figure \ref{fig:scatter} with orange color. 

However, we find that the fitting algorithm tends to underestimate the $\nu_F$ for those with the extended profiles, which correspond to more abrupt rise of power in the central region of the Fourier space, and therefore a worse fitting. Consequently, selection based purely on $\nu_{F,fit}$ would tend to discard galaxies of more extended profiles. To avoid this problem while keeping the advantage of $\nu_{F,fit}$, we propose to use the maximum between $\nu_F$ and $\nu_{F,fit}$ (called $\widetilde{\nu}_F$ hereafter) as our proposed selection criterion for Fourier\_Quad instead of $\nu_F$. We find that our results and conclusions in the previous sections are hardly affected by this change. 

\subsection{Weight for the Fourier\_Quad method}
Note that we estimate the shear signal by taking the ensemble averages of the shear estimators (Eq.(\ref{shear_measurement})). However, the amplitudes of the shear estimators defined in Eq.(\ref{shear_estimator}\&\ref{TM}) are proportional to the square of the galaxy flux, and the ensemble averages would be dominated by the bright galaxies if the shear estimators are not weighted. In the results presented so far, we weight each shear estimator using $F^{-2}$, with $F$ being the true galaxy flux. Note that as the true flux does not correlate with the underlying shear signal, it should not introduce any weight-related bias. Since our newly defined $\tilde{\nu}_F$, which is proportional to the square of the galaxy flux, does not correlate with the galaxy shape, it can also be used to weight the shear estimators without introducing systematic biases. 

\begin{figure}
	\centering
	\includegraphics[width=0.4\textwidth]{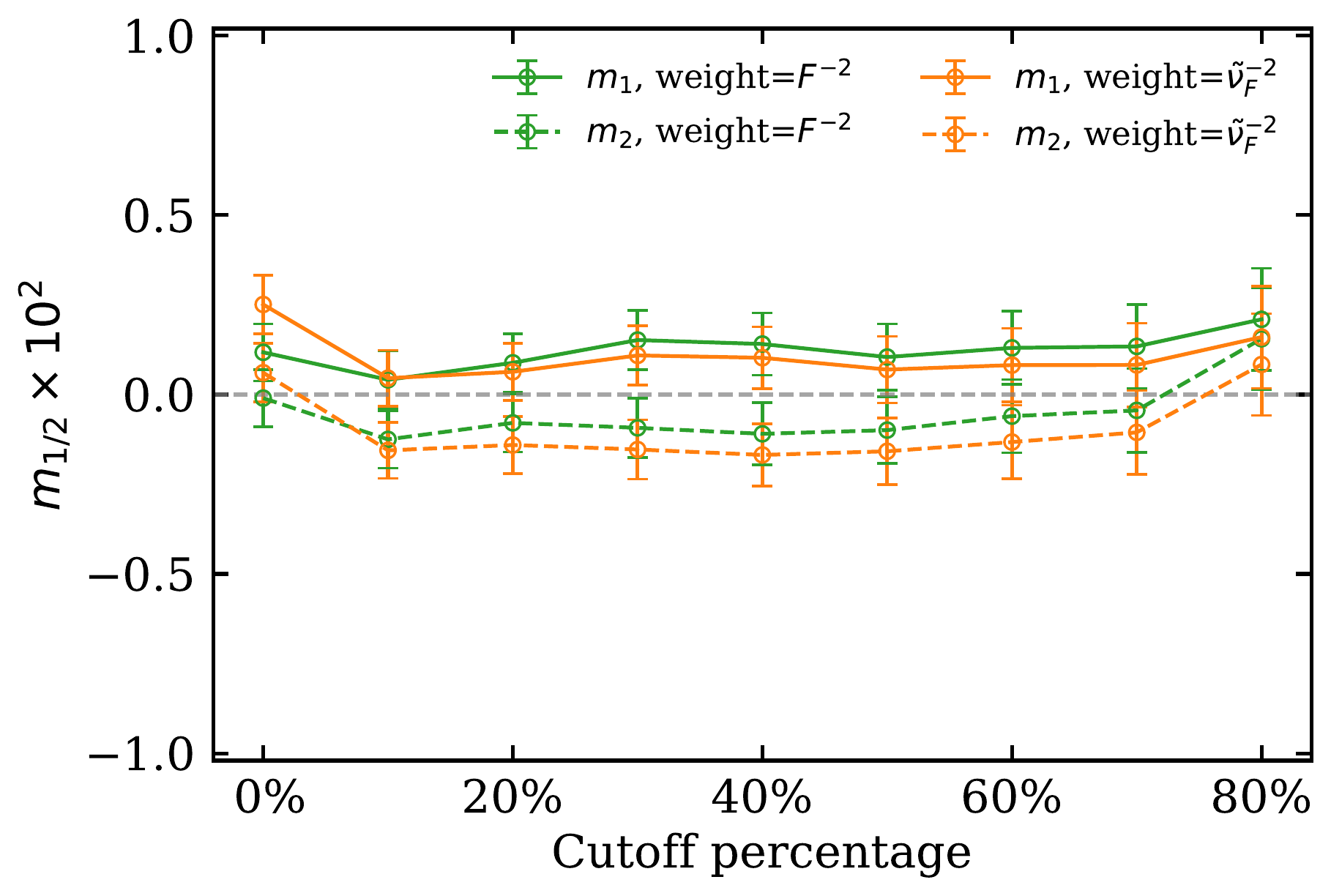}
	\caption{Comparison of the multiplicative biases for shear estimators weighted by the true flux $F$ and $\widetilde{\nu}_F$ respectively using the PI sample. The results are similar for other samples.\label{fig:weight_compare}}
\end{figure}

In practice, as $F$ is not available, we find that our new selection criterion $\tilde{\nu}_F$ is a qualified replacement of $F$. Figure \ref{fig:weight_compare} shows the measurement of the multiplicative biases for the PI sample. For the green curves in the figure, $\tilde{\nu}_F$ is used not only for selecting the sample, but also as a weight function. It shows that neither the selection nor the weighting introduce any noticeable shear bias in this case. This is perhaps not surprising, as selection is essentially a type of weighting, and they rely on the same mechanism (correlation with the galaxy shape) in generating the shear bias.   

\section{Conclusion}
\label{conclusion}

Sample selection may introduce shear bias if the selection criterion is correlated with the galaxy shape. In this paper, with the Fourier\_Quad shear measurement method, we study the performance of several selection criteria, including magnitude, SNR, and resolution factor, as well as a new function $\nu_F$, which is defined as a measure of the signal-to-noise ratio within a fixed domain in the neighborhood of the galaxy. The selection bias is measured on simulated galaxies with both regular shapes (generated by Galsim) and irregular ones (made of point sources connected by random walks).

The selection effect can introduce both multiplicative and additive bias (when the PSF has an anisotropic form) as a result of the couplings between the selection criteria and the galaxy shape. This is shown in \S\ref{sec:m} and \S\ref{sec:a} with both analytical arguments and numerical evidences. We find that all three traditional selection criteria have non-negligible sensitivities to galaxy shapes, leading to a multiplicative shear bias at the level of a few percent, and an additive bias proportional to the PSF ellipticity. Selections according to the magnitude introduce multiplicative bias at the level of $\sim 1-2\%$, and those according to SNR or the resolution factor can cause much a larger multiplicative bias. In general, the bias is larger on irregular galaxies than on the regular ones. 

In contrast, our newly defined selection criterion $\nu_F$ performs much better. It works almost as well as the true magnitude (pre-lensing). $\nu_F$ uses the power at $\vec{k}=0$ in Fourier space as an estimator of the galaxy flux. In this case, the domain for counting the galaxy flux is fixed. $\nu_F$ is therefore much less sensitive to the galaxy shape than the other popular selection criteria (Figure \ref{fig:change}), and can be safely used as a bias-free selection criterion in shear measurements. We also propose $\widetilde{\nu}_F$ as a slightly modified version of $\nu_F$ to reduce the scatter at the faint end, without changing its quality as a selection criterion. When the ensemble averages are taken for the shear estimators of Fourier\_Quad, $\widetilde{\nu}_F$ can also be used as a bias-free weighting function for homogenizing the contributions from galaxies of different luminosities. 

In our current study, detection-related selection effect is not considered, as the detection rate of our simulated sample is very high (according to the result of SExtractor). In practice, to avoid detection-related shear bias, one can set the threshold of the selection criterion high enough. But we caution that there are other source selection effects that can bias the shear measurements, such as image overlapping \citep{Sheldon2019} and photo-z error. We plan to study these effects in a future work with real data. We also plan to extend our current discussions to issues related to the PDF\_SYM approach, which is a promising new statistical approach in Fourier\_Quad method. 

The new selection criterion $\tilde{\nu}_F$ has been used in our Fourier\_Quad pipeline\footnote{The software is available upon request.} to deal with the CFHTLenS data. Part of the early results have been presented in \cite{Zhang2019}. We believe that $\tilde{\nu}_F$ is also useful for other shear measurement methods, because our discussion in the section \ref{sec:m} \& \ref{sec:a} regarding the shear sensitivity of the selection criterion is independent of shear measurement. A detailed discussion of this topic is however beyond the scope of this paper.

\acknowledgments
This work is supported by the National Key Basic Research and Development Program of China (No.2018YFA0404504), and the NSFC grants (11621303, 11890691, 11673016). Dezi Liu acknowledges the support of the Launching Research Fund for Postdoctoral Fellow from the Yunnan University with grant C176220200 and the China Postdoctoral Science Foundation with grant 2019M663582. Jiajun Zhang was supported by IBS under the project code, IBS-R018-D1. The computations in this paper were run on the $\pi$ 2.0 cluster supported by the Center for High Performance Computing at Shanghai Jiao Tong University.

\end{document}